\documentclass{article}
\usepackage{spconf,amsmath,graphicx}

\usepackage{enumitem}

\usepackage{hyperref}
\usepackage{epsfig}
\usepackage{graphicx}
\usepackage{url}
\usepackage{color}
\usepackage{mathrsfs}  
\usepackage{amssymb,amsmath,bm}
\usepackage{multicap,graphicx}
\usepackage{caption}
\usepackage{algorithm}  
\usepackage{algorithmic}
\usepackage{graphicx}
\usepackage{wrapfig}
\usepackage{picinpar}
\usepackage{cutwin}
\usepackage{stmaryrd}

\title{Lung segmentation with NASNet-Large-Decoder Net}
\name{Youshan Zhang \ \ \ \ \ }
\address{Computer Science and Engineering, \\Lehigh University, Bethlehem, PA, USA \\ yoz217@lehigh.edu}

\begin{document}
%
\maketitle
\begin{abstract}
Lung cancer has emerged as a severe disease that threatens human life and health. The precise segmentation of lung regions is a crucial prerequisite for localizing tumors, which can provide accurate information for lung image analysis. In this work, we first propose a lung image segmentation model using the NASNet-Large as an encoder and then followed by a decoder architecture, which is one of the most commonly used architectures in deep learning for image segmentation. The proposed NASNet-Large-decoder architecture can extract high-level information and expand the feature map to recover the segmentation map. To further improve the segmentation results, we propose a post-processing layer to remove the irrelevant portion of the segmentation map. Experimental results show that an accurate segmentation model with 0.92 dice scores outperforms state-of-the-art performance.\footnote{This work is written before our previous paper~\cite{zhang2021automatic}.}
\end{abstract}

\section{Introduction}

Lung cancer is one of the malignant tumors with the fastest increase in morbidity and mortality and the greatest threat to human health and life. In the past fifty years, many countries have reported a marked increase in the incidence and mortality of lung cancer \cite{alberg2013epidemiology}. The lung-related disease is kept on the list of top ten causes of death in the country in the United States ~\cite{Rui2015National,murphy2017deaths}. Therefore, the prevention of lung cancer is important.

From the definition, lung cancer is a malignant lung tumor that is characterized by uncontrollable growth in the lung tissue. Clinicians find that the cure rate of early lung cancer is over 90\%. Hence, the early diagnosis of lung cancer is essential since it could control the disease, reduce the mortality rate, and increase the patient's survival rate when the treatment is more likely curative. Chest radiograph (CXR) is commonly performed as diagnostic imaging for lung cancer and Pneumonia diagnosis. However, there are a number of factors, such as the positioning of the patient and depth of inspiration, that can affect the appearance of the CXR, complicating interpretation further. In addition, clinicians are faced with reading high volumes of images every shift. It is difficult for the physician to obtain an accurate diagnosis without the help of an additional tool. Therefore, it is necessary to develop segmentation tools to improve the effectiveness of the treatment.

However, designing an effective lung segmentation method is a challenging problem since the ROIs are often confused with the lung tissue.
A large number of medical image analysis techniques have been proposed for lung segmentation, such as threshold method \cite{hu2001automatic,pu2008adaptive,leader2003automated}, region-based method \cite{ray2003merging,guorong2013multi}, genetic method\cite{ozekes2007rule}, level set method \cite{vese2002multiphase,silveira2007automatic,farag2013novel} and artificial neural network \cite{lin2005autonomous,ceylan2010novel,kuruvilla2014lung}, etc.

The threshold method is one of the most common and straightforward segmentation methods in lung segmentation. It is a region segmentation technology, which divides the gray value into two or more gray intervals, chooses one or more appropriate thresholds to judge whether the region meets the threshold requirement according to the difference between the target and the background, and separates the background and the target to produce a binary image. Threshold processing has two forms: global threshold and adaptive threshold. The global threshold only sets one threshold, and the adaptive threshold sets multiple thresholds. The target and background regions are segmented by determining the threshold at the peak and valley of the gray histogram \cite{hu2001automatic,pu2008adaptive}. Level set methods are also widely used in the segmentation task. The basic idea of the level set method for image segmentation is the continuous evolution of curve motion. The boundary of the image is searched until the target contour is found and then the moving curve is stopped. Curves are moved along every three-dimensional section of images to slice different levels of the three-dimensional surface. The level of the obtained closed curves of each layer change over time, and finally get a corresponding shape extraction contour \cite{vese2002multiphase}.

\begin{figure*}[h]
\centering
\includegraphics[width=2 \columnwidth]{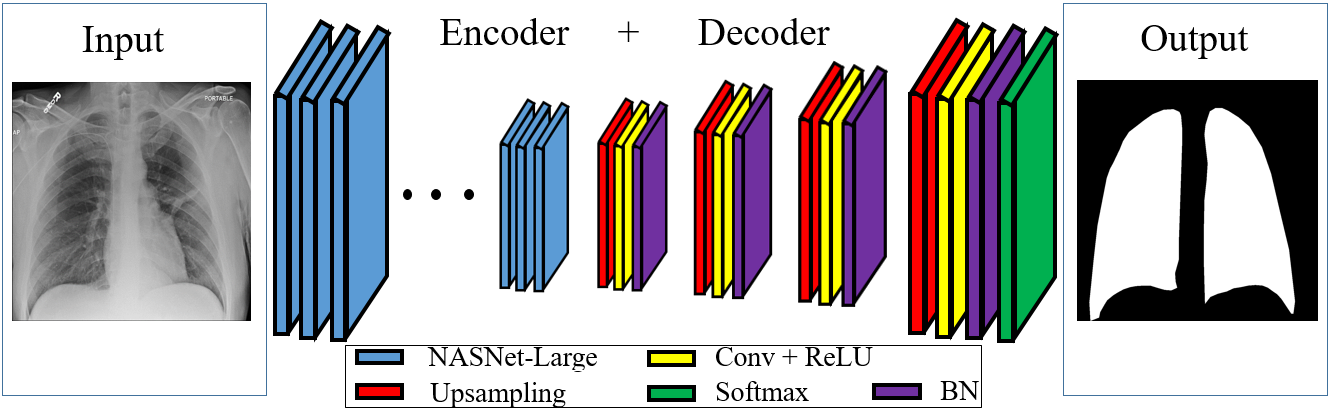}
\caption{The architecture of NASNet-Large segmentation net. The encoder consists of the first 414 layers from the NASNet-Large model. There are four blocks in the decoder, and each block contains Upsamling, Conv+ReLu, and BN layer. (Convolution (Conv), Batch normalization (BN), Rectified linear units (ReLU). The final decoder is fed into a softmax layer for pixel-wise lung prediction.   }
\label{fig:Nastnet_segnet}
\end{figure*}
Deep neural networks have also been applied in segmentation, and they supersede many traditional image segmentation approaches. Garcia published a review of deep learning approaches that aimed to present an overview of deep learning-based segmentation \cite{garcia2017review}. There are several models to address the segmentation task. Fully convolutional network (FCN) \cite{long2015fully} is one of the earliest works for deep learning-based image segmentation problem, which performs end-to-end segmentation. It is a convolution network for a dense prediction that does not need to pass through the full connection layer. This method leads to the possibility of segmenting images of any size effectively, and it is much faster than the patch classification method. Almost all of the other more advanced methods follow this architecture. However, there are several limitations of the FCN model, such as the inherent spatial invariance causes the model fails to take into account of useful global context information, and its efficiency in high-resolution scenarios is worse and is not available for real-time segmentation. Another problem that causes difficulty in using CNN networks in segmentation is the existence of pooling layers. The pooling layer not only enlarges the sensing field of the upper convolution layer but also aggregates the background and discards part of the location information. However, the semantic segmentation method needs to adjust the category map accurately, so it requires retaining the location information abandoned in the pooling layer. Later encoder-decoder architecture is widely used in the segmentation. Segnet \cite{badrinarayanan2017segnet} and U-net \cite{ronneberger2015u} are representative encoder-decoder architectures. It first selects a classification network such as VGG-16 and then removes its fully connection layer to produce low-resolution image representation or feature mapping. This part of the segmentation network is usually called an encoder. A decoder is part of the network, which can learn how to decode or map these low-resolution images to the prediction at the pixel level. The difference between different encoder-decoder architectures is the design of the decoder. Another type of method uses dilated convolutions layers and removes the pooling layer structure. Chen et al. proposed the Deeplab model, which used the dilated convolutions, and it used the fully connected conditional random field to realize the atrous spatial pyramid pooling (ASPP)  in the spatial space \cite{chen2014semantic}.

In this paper, we provide a lung segmentation model using one of the common encoder-decoder architectures for image segmentation with a deep learning model called NASNet-Large-decoder. This architecture can erase unnecessary information provided in lung images. Our network
achieves an accurate segmentation result with a 0.92 dice score.

Our contributions are in three folds:
\begin{enumerate}
    \item We are the first to define a NASNet-Large-decoder net to segment the chest radiography images;
    \item To remove the irrelevant small segmented parts, we are the first to propose a post-processing layer, which is able to filter out the false segmented section in prediction images; 
    \item  Experiments results demonstrate that the proposed model achieves the highest dice and IoU score over state-of-the-art.
\end{enumerate}

This paper is organized as follows: in section~\ref{sec:method}, the NASNet-Large-decoder net architecture is summarized; we present the segmentation results in section~\ref{sec:results}; In section ~\ref{sec:dis}, we discuss the advantages and disadvantage of the proposed model and conclude in section~\ref{sec:conclusion}.

\section{Method}\label{sec:method}

In this section, we first introduce the proposed segmentation net and then describe a post-processing layer for filtering out the irrelevant segmented section in the prediction image.  

\subsection{NASNet-Large segmentation net}
The NASNet-Large segmentation net contains an encoder and decoder, which is followed by a classification layer. The architecture is shown in Fig.\ref{fig:Nastnet_segnet}. There are two significant differences in our model compared with Segnet, which employs the pre-trained vgg16 network for the encoder. Our NasnetLarge-decoder net uses the first 414 layers of NasnetLarge net (which is a well-trained classification net on ImagNet) as the encoder to decompose images \cite{zoph2018learning}. We do not use the pre-trained weighted but retrain the layers using the new data to fit the NasnetLarge net in our experiment since our dataset is significantly different from the ImageNet. 
Another one is that the decoder is different, and there are no pooling indices in our model since the NasnetLarge net can produce detailed information for the decoder. 

An appropriate decoder in the decoder network can upsample its input feature map using the max-pooling layer. The decoding technique is illustrated in Fig~\ref{fig:Nastnet_segnet}. There are four blocks in the decoder. Each block first begins with an upsampling layer, which can expand the feature map, and then feature maps are followed by convolution and rectified linear units. A batch normalization layer is then applied to each of these maps. The first decoder, which is closest to the last encoder, can produce a multi-channel feature map. This is similar to the Segnet, which can generate a different number of sizes and channels as their encoder inputs. 
The final output of the last decoder is fed to a trainable soft-max classifier. And the output of this softmax layer is a $K$ channel image of probabilities where $K$ is the number of classes (two in our problem). The predicted segmentation corresponds to the class with maximum probability at each pixel.

\subsection{Post-processing layer }
\begin{figure}[t]
\centering
\includegraphics[width=0.95 \columnwidth]{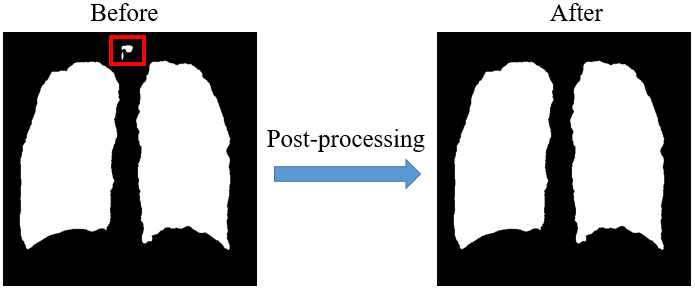}
\caption{The post-processing of prediction result. The left one is the prediction result from the proposed net, and the right one is the post-processing result using the proposed post-processing layer. The red box is an irrelevant feature.}
\label{fig:post}
\end{figure}

However, there are some small parts that are not the true lung in the prediction result. We then propose a post-processing layer, which can filter the irrelevant part in the image. The first step of the post-processing layer is to classify the predicted image into several parts\footnote{The major step of the post-processing step is using the connectedComponentsWithStats function in OpenCV. }, and we then select the first two largest areas (the lung area) as the final segmented lung. As shown in the left image in Fig.\ref{fig:post}, the red box is the wrong prediction of the image (false negative). After the post-processing, the red box is removed. Therefore, the prediction result will be improved if we filter out these irrelevant parts in images. 

\section{Results}\label{sec:results}
\subsection{Datasets and parameters}
\begin{figure}[t]
\centering
\includegraphics[width=0.95 \columnwidth]{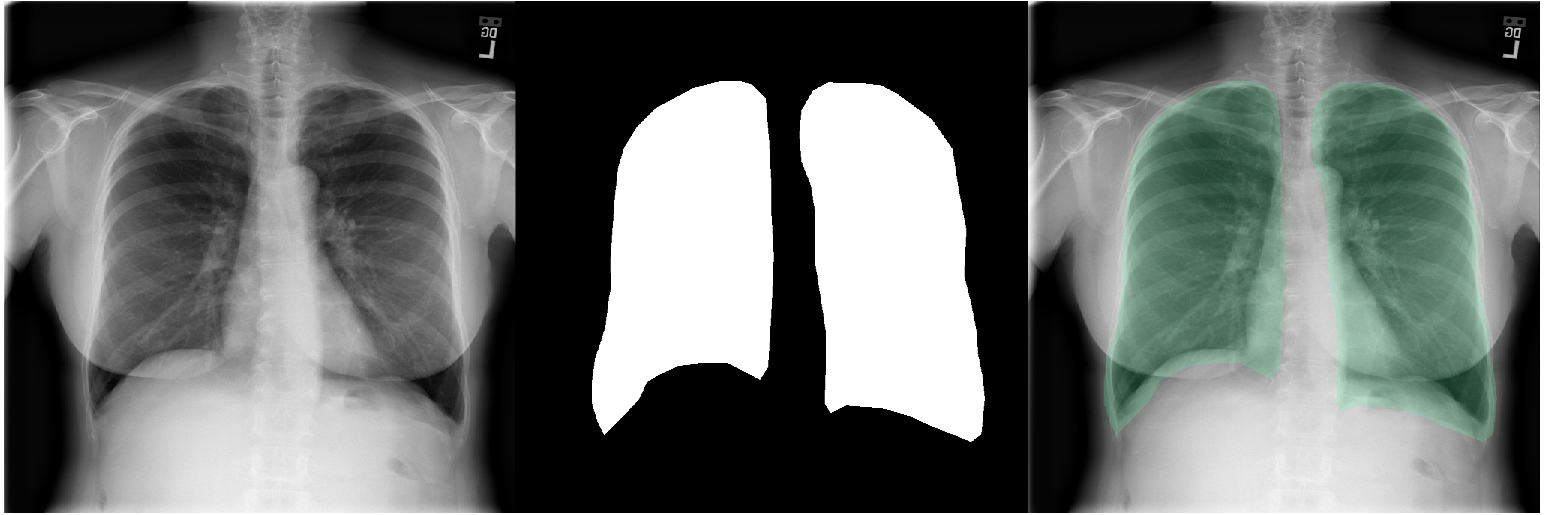}
\caption{An example lung image in the training dataset (left: raw image; middle: ground truth mask image; right: mask overlay with raw image).}
\label{fig:exam}
\end{figure}

Our lung segmentation data is from the RSNA pneumonia detection dataset. The whole dataset can be downloaded from \url{https://www.kaggle.com/c/rsna-pneumonia-\\detection-challenge}. To remove the unrelated features, we focus on lung segmentation; the lung is manually segmented\footnote{The dataset is available at: \url{https://github.com/YoushanZhang/Lung_Segmentation}.}. There are 800 training images and 200 test images in this dataset. Fig.~\ref{fig:exam} shows an example of a lung image in the training dataset.

Our experimentation is based on Keras, which is a high-level neural networks API written in python and is able to run on top of either TensorFlow or Theano and runs seamlessly on CPU and GPU. In addition, our network was trained on a graphics processor NVIDIA TITAN Xp equipped with 12Gb of memory in order to exploit its computational speed. Hence, we run our code on a GPU in order to greatly accelerate the execution with the TensorFlow backend. The network parameters are set to:

\begin{enumerate}
    \item Batch size: 4
    \item Step  size: 5
    \item Number of epochs: 100
\end{enumerate}

\subsection{Metrics}
To evaluate the performance of our  NASNet-Large segmentation net, we use the dice coefficient index as a similarity metric to indicate the goodness of the segmentation results since the dice score is currently widely used in the segmentation task. Furthermore, we also report the IoU score. The two metrics are defined in the following formula:

\begin{equation*}
    Dice=2 \times \frac{|A \cap  B|}{|A|+|B|}, \ \ \ IoU=\frac{|A \cap B|}{|A \cup B|},
\end{equation*}
where $A$ is ground truth mask, and $B$ is the prediction mask.

\begin{figure}[t]
\centering
\includegraphics[width=1 \columnwidth]{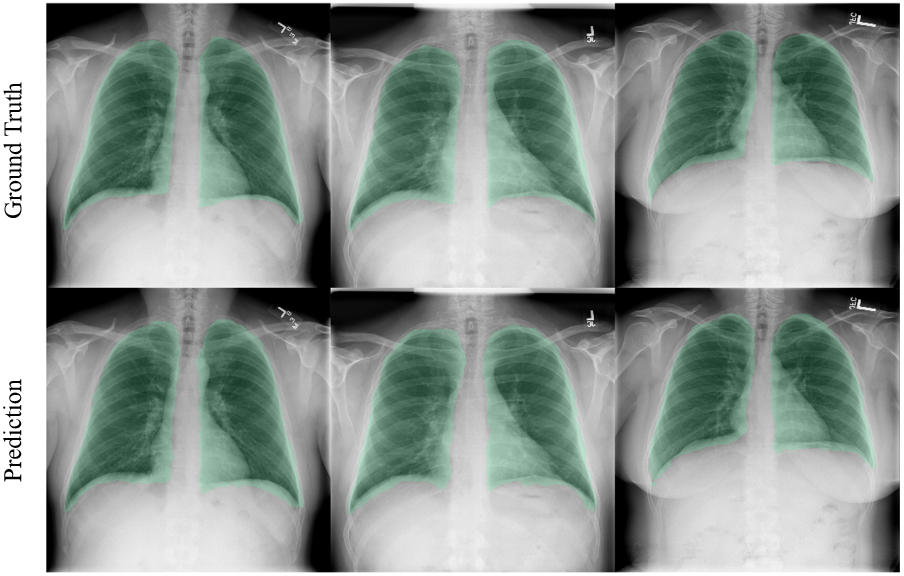}
\caption{Ground truth VS. prediction results. The first row is the ground truth mask overlay with the raw image, and the second row is the prediction result overlay with the raw image.}
\label{fig:com}
\end{figure}

\begin{table}[h]
\small
\begin{center} 
\caption{Segmentation results comparison}
 \setlength{\tabcolsep}{+6mm}{
\begin{tabular}{|c|cc|c|c|c|c|c|c|c|c|c|}
\hline \label{tab:R2}
Methods &  IoU & Dice score \\
\hline
Segnet  \cite{badrinarayanan2017segnet}& 0.82  & 0.87 \\
U-Net \cite{ronneberger2015u} & 0.84  & 0.88 \\
Deep-Lab \cite{chen2014semantic}  & 0.85  & 0.89 \\
\hline
\hline
{\bf Nastnetlarge-net}   & \textbf{0.86}  & \textbf{0.91} \\
{\bf Nastnetlarge-net-Post}   & \textbf{0.87}  & \textbf{0.92} \\
\hline
\end{tabular}}
\end{center}
\end{table} 

\subsection{Segmentation results}
As shown in Fig.\ref{fig:com}, it compares the predicted segmented lung image with the ground truth image in the test dataset. The prediction image is close to the real mask, which demonstrates the high performance of our model. We also compare prediction results with state-of-the-art methods. Tab.~\ref{tab:R2} lists the comparison results of different models (Notice that the metrics are only reported on the lung area and exclude the background area). Our Nastnetlarge-net has a higher IoU and dice score than other models, and we also observe that the post-processing layer outputs the highest scores, which illustrates that the post-processing layer is useful in the segmentation tasks.

\section{Discussion}\label{sec:dis}
\begin{figure}[t]
\centering
\includegraphics[width=1 \columnwidth]{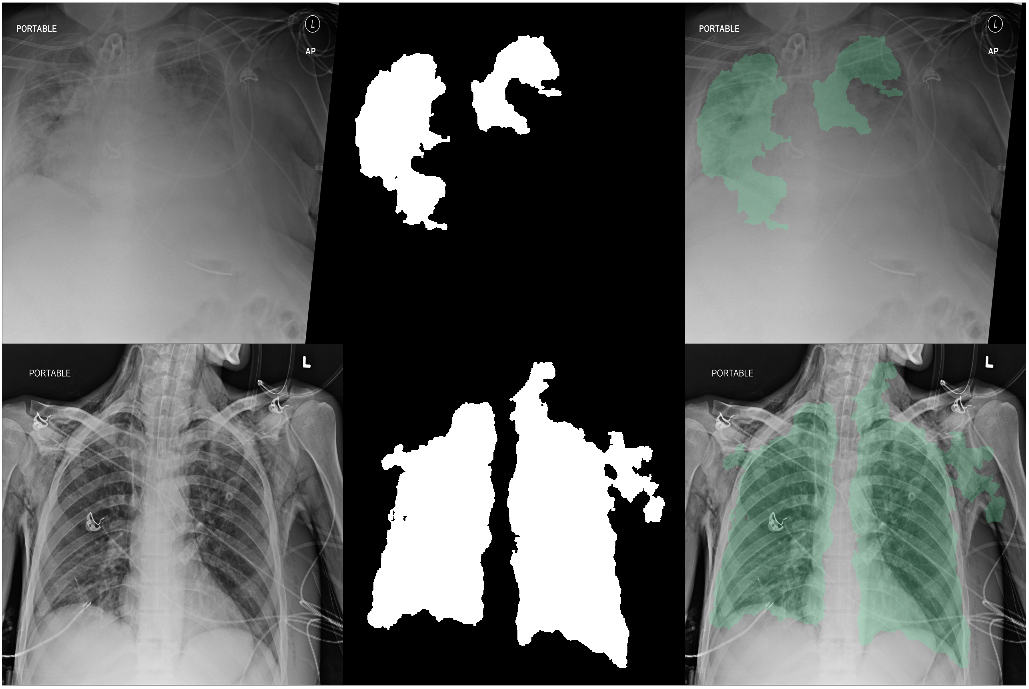}
\caption{Two failures in our model. The first row is caused by the light parts, and the second row is caused by the dark parts.}
\label{fig:bad}
\end{figure}

One of the distinct advantages of our model is that it achieves a higher dice and IoU score. And there are two reasons: the designed NASNet-Large segmentation net is suitable for image segmentation, and the post-processing layer filters out the unnecessary parts in the image, which improves the segmented results.
Although our model achieves a 0.92 dice score, it still fails in some cases. As shown in Fig.~\ref{fig:bad}, we observe that the segmentation results are worse in these two situations. There are also two reasons, and one is that there is no similar image in the training dataset, which leads to low performance. Second is that our model is not robust enough to deal with some unusual images. Compared with the normal training images as shown in Fig.~\ref{fig:exam}, two worse raw images (in Fig.~\ref{fig:bad}) are either too light (first row) or too dark (second row), and it is even difficult for a human to distinguish the lung area and other tissues. A straightforward solution is to include more similar images in the training images. 

\section{Conclusion} \label{sec:conclusion}
In this paper, we are the first to present a lung segmentation using the NASNet-Large-decoder architecture, and we get an accurate
segmentation with a 0.92 dice score. A post-processing layer is employed to remove the unnecessary part of the prediction map. The proposed model can also be applied to a wide area of different medical image segmentation tasks. Our objective in the next stage is to design a more robust encoder and decoder model for application in all different cases.

\small
\bibliographystyle{IEEEbib}
\bibliography{strings,refs}

\end{document}